\documentclass[10pt,letterpaper]{article}
\usepackage[top=0.85in,left=2.75in,footskip=0.75in]{geometry}

\usepackage{amsmath,amssymb}

\usepackage{changepage}

\usepackage[utf8x]{inputenc}

\usepackage{textcomp,marvosym}

\usepackage{cite}

\usepackage{nameref,hyperref}

\usepackage[right]{lineno}

\usepackage{microtype}
\DisableLigatures[f]{encoding = *, family = * }

\usepackage[table]{xcolor}

\usepackage{array}

\newcolumntype{+}{!{\vrule width 2pt}}

\newlength\savedwidth

\usepackage{tikz}
\usepackage{graphicx}
\usetikzlibrary{arrows.meta}

\usepackage{setspace} 

\raggedright
\usepackage[aboveskip=1pt,labelfont=bf,labelsep=period,justification=raggedright,singlelinecheck=off]{caption}

\makeatletter
\renewcommand{\@biblabel}[1]{\quad#1.}
\makeatother

\date{}

\usepackage{lastpage,fancyhdr,graphicx}
\usepackage{epstopdf}
\fancyheadoffset[L]{2.25in}
\fancyfootoffset[L]{2.25in}
\begin{document}
\vspace*{0.2in}

\begin{flushleft}
{\Large
\textbf\newline{Bourdieu, networks, and movements: Using the concepts of habitus, field and capital to understand a network analysis of gender differences in undergraduate physics} 
}
\newline
\\
Steven Martin Turnbull\textsuperscript{1,2,\textcurrency,*},
Fr\'ed\'erique Vanholsbeeck \textsuperscript{3,4},
Kirsten Locke\textsuperscript{1}
Dion R. J. O'Neale\textsuperscript{2,3},
\\
\bigskip
\textbf{1} Critical Studies in Education, Faculty of Education and Social Work, University of Auckland, Auckland, New Zealand
\\
\textbf{2} Te P\={u}naha Matatini, Auckland, New Zealand
\\
\textbf{3} Department of Physics, University of Auckland, Auckland, New Zealand
\\
\textbf{4} The Dodd-Walls Centre, Auckland, New Zealand

\bigskip

%
%


\textcurrency Current Address: The Department of Physics, The University of Auckland, Private Bag 92019, Auckland, New Zealand 

* s.turnbull@auckland.ac.nz

\end{flushleft}
\section*{Abstract}
Current trends suggest that significant gender disparities exist within Science, Technology, Engineering, and Mathematics (STEM) education at university, with female students being underrepresented in physics, but more equally represented in life sciences (e.g., biology, medicine). To understand these trends, it is important to consider the context in which students make decisions about which university courses to enroll in. The current study seeks to investigate gender differences in STEM through a unique approach that combines network analysis of student enrollment data with an interpretive lens based on the sociological theory of Pierre Bourdieu. We generate a network of courses taken by around 9000 undergraduate physics students (from 2009 to 2014) to quantify Bourdieu's concept of field. We explore the properties of this network to investigate gender differences in transverse movements (between different academic fields) and vertical movements (changes in students' achievement rankings within a field). Our findings indicate that female students are more likely to make transverse movements into life science fields. We also find that university physics does a poor job in attracting high achieving students, and especially high achieving female students. Of the students who do choose to study physics, low achieving female students are less likely to continue than their male counterparts. The results and implications are discussed in the context of Bourdieu's theory, and previous research. We argue that in order to remove constraints on female student's study choices, the field of physics needs to provide a culture in which all students feel like they belong.


\section*{Introduction}
Historically, women have been underrepresented in Science, Technology, Engineering and Mathematics (STEM) disciplines. This is a concerning issue today internationally, and at all stages of higher education \cite{Abraham_2014,Stevanovic_2013,Smith_2011}. More recent studies indicate specific gender disparities exist within the sub-fields that comprise STEM \cite{mullis2016timss}. Female students tend to be underrepresented in physics in higher education, and this is evidenced by research from the United States \cite{NSF, Cunningham_2015,Kost_Smith_2010,Heilbronner_2012}, Europe \cite{Huyer2007, Stevanovic_2013, InstituteofPhysics_2012, InstituteofPhysics_2013, Smith_2011}, Asia-Pacific regions \cite{EducationCounts_2016a, Kennedy_2014} and Africa \cite{Semela_2010}. In contrast, the same research shows that the life science subjects (biology and medicine) tend to have more of a gender balance. Why do we see gender differences in the physical and mathematical science subjects, but not the life science subjects? Much research has been dedicated to understanding the extent, causes, and possible solutions to this issue \cite{Brewe_2016, Blickenstaff_2005}. 

The current study investigates the outcomes for male and female physics students at the University of Auckland (UoA) --- the largest university in New Zealand. We adopt a unique approach, by combining quantitative network analysis with a research framework based on Pierre Bourdieu's sociological theory. Whilst we argue that these two approaches can provide a detailed understanding of gender disparities in student enrolment patterns, there is a lack of research in this area (for examples of how network analysis and Bourdieu have been previously used together ,see the work of de Nooy \cite{de2003fields}), and Bottero and Crossley \cite{bottero2011worlds}). We combine these approaches by using network analysis to provide a representation of Bourdieu's concept of field, with an emphasis on his ideas of transverse and vertical movements (students moving from one field to another, and moving upwards and downwards in achievement rankings in a field). In order to avoid misinterpretation of Bourdieu's theory, which is easily done when ``bits and pieces'' of it are used \cite[p.4]{Bourdieu1992}, we combine our representation of field with Bourdieu's concepts of habitus and capital. We argue that network analysis can bring to light the complex patterns of students' subject enrollment, whilst Bourdieu's theory offers a rich theortical framework to explain these patterns. We place the findings of our network analysis in a broad socio-cultural context that brings to light the complex interactions between society, gender and subject discipline. To avoid confusion, the following sections will use `field' as a technical term referring to the Bourdieu's definition (which will be explained in more detail in the next section), and `discipline' as a non-technical term that describes the different STEM domains. 

We begin by introducing a simple model of Bourdieu's theory, using the field of science education to illustrate its concepts. We then add to this outline of theory by building our method of network analysis into Bourdieu's theory. More specifically, we describe how network analysis of student enrollment data can provide a representation of field. Exploring the properties of this network structure allows us to understand gender differences in the movements students make within and across fields. According to Bourdieu \cite[p.131]{Bourdieu1984}: 
\begin{quote}
    The social space, being structured in two dimensions (overall capital volume and dominant/dominated capital) allows two types of movement... vertical movements, upwards or downwards in the same vertical sector, that is in the same field... and transverse movements, from one field to another, which may occur either horizontally or between different levels.
\end{quote}In science education, individual's may move from one field to another (i.e., from physics to life science), but also upwards and downwards in achievement rankings in the field. We use these concepts of movements to guide our investigation. We seek to understand whether there are gender differences in the number of students moving from physics to other fields, and also in the changes in achievement rankings of students in physics. We close this article with a discussion of our results in the broader context of previous research and Bourdieu's concepts of capital and habitus. 

\section{Theoretical Framework}
The metaphor of the leaky pipeline is often used to describe the attrition of women from physics \cite{Huyer2007, Schiebinger_2001}, in that women are more likely to drop out with each transition between key stages of education (particularly secondary school to university). This metaphor can be criticized for not only stigmatising individuals that drop out of the pipeline, but for also being too simplistic \cite{Cannady2014}. It is important to emphasize contextual factors, such as the presence of gender-stereotypes \cite{Nosek_2009} (e.g., men study science, women study humanities) that impact on the decisions that students make. It is also important to consider the complex nature of students' enrollment patterns; in reality a student's journey through university study follows a complex network of unique pipes, rather than a singular pipeline. The current study employs a research framework that builds on the limitations of the leaky pipeline. We seek to place our results in a wider socio-cultural context, by harnessing a research framework adapted from the work of Pierre Bourdieu \cite{Bourdieu1984} (see Fig~\ref{fig:Fig1}). We employ Bourdieu's concepts of capital, habitus and field to interpret our findings, and place them in the context of previous studies that have investigated gender differences in STEM subject selection. 

The following sections will outline Bourdieu's concepts of field, capital and habitus. We apply these concepts to a host of previous research regarding gender disparities in science to outline the socio-cultural context in which students are placed. More specifically, we outline research that describes the state of the field of physics and the distribution of capital within the field of physics. We then discuss the interaction of capital with habitus --- the system of dispositions that is formed in relation to the field. We describe how the ``smog of bias'' \cite[p.1]{Kost_Smith_2010} that targets women in physics may impact on habitus, and thus practices within the field, such as choosing to discontinue physics study. 

\subsection{Field}
For Bourdieu, the world is separated into a collection of different fields\cite{Bourdieu1984}. A field can be considered as a system of social locations, where each individual is objectively ranked by the resources (capital) they have relative to others. For example, in the field of tertiary science education, a lecturer ranks higher than a student, whilst a high achieving student ranks higher than a low achieving student. To begin to the understand the hierarchical nature of a field, we must first understand the concept of \textit{capital}. Originally conceived within economics, capital was defined by Adam Smith (in 1887) as ``That part [of a person's wealth] that he expects to provide [them] with \textellipsis income\textellipsis'' \cite[p.214]{Smith_1887}. Bourdieu interpreted capital as a legitimate, valuable and exchangeable resource that individuals can use to gain advantage in society \cite{Bourdieu_1986}. 
Therefore, the rankings are determined by how we define what is valuable and legitimate in the field. The practices of individuals within the field are judged by criteria internal to the domain of activity \cite{hilgers2014introduction}. Individuals with a high volume of valued capital will hold power within the field. For example, high achieving students have high volumes of capital in the field due to their course grades (a signal of success), whilst lecturers and researchers have a greater volume of capital in the form of qualifications and research experience. In the field of tertiary science education, lecturers and researchers sit at the top of the hierarchy, and decide what kinds of capital are valued or devalued (e.g., professors often decide the course content and manner of teaching for undergraduate students at university). We will discuss Bourdieu's conceptualization of capital and the way it can inform gender equity research in the following section. Before then, we will outline a brief description of how the field of physics is structured from an objective point of view in relation to gender.

The numbers of male and female students holding qualification in the different science disciplines can provide an objective, surface level understanding of the structure of the field. In the United States, only around 20\% of students studying physics at bachelors, masters or doctorate level in 2014 were female\cite{NSF}. This contrasts with biology, where around 50-60\% of students studying at bachelors, masters or doctorate level were female\cite{NSF}. Similar gender disparities in physics enrollments have been found in the European \cite{InstituteofPhysics_2012,InstituteofPhysics_2013, Stevanovic_2013}, and Asia-Pacific regions \cite{Abraham_2014, Kennedy_2014}. Data from UNESCO  shows that, in Europe in 2007, around 71\% of tertiary health and welfare students were female, whilst this figure was 39\% for natural and physical science (biology, physics, chemistry) \cite{Huyer2007}. 

Reports from New Zealand in 2017 show that, overall, secondary school science had a balanced gender-ratio of year 13 (i.e. final year of high-school) students \cite{EducationCounts_2016a}. However, male students dominated physics and mathematics from year 11 to year 13 at secondary school, with this trend being reflected at university level \cite{EducationCounts_2016a}. Across the same school years, biology and human anatomy tended to have more female students than male students. Looking at tertiary science education (i.e. university undergraduate and post-graduate levels) these gender disparities were maintained \cite{EducationCounts_2016a}. Other data from New Zealand in 2017 shows that female students were slightly less well represented among bachelor students studying physics and mathematics (43\% and 46\% respectively) \cite{EducationCounts_2016b}. At the same level, female students tended to be over-represented in biology and health (67\% and 74\% respectively) \cite{EducationCounts_2016b}. Approximately 25\% of doctoral students in physics and astronomy and 44\% of students in mathematics were female, while female students comprised 53\% of students studying biology and 69\% of those studying health. Beyond post-graduate level study, the representation of women in New Zealand professorial roles and leadership positions in physics is particularly poor \cite{Bray_2011}. 

The above outlines clear evidence of gender disparities in the field of science, internationally and in New Zealand specifically. Whilst useful, these figures only provide a static, surface-level understanding of what is happening in the field of science education. As shown in Model \ref{fig:Fig1}, the generation of the practices and behaviours represented in the field (such as enrollment patterns) are generated through the interaction of capital and habitus. We will now visit these two concepts, applying them to previous research, to understand why gender disparities in science education are common.   

\subsection{Capital}
The objective rankings within a field are defined by the distribution of capital, and this can be used to inform gender equity research in science \cite{Kelly_1985}. Different fields have different forms of logic as to what forms of capital are of value. Using a basic example, a science qualification is worth more in the field of science than in other academic fields. Capital is complex and may take many forms, each of which may be valued differently depending on the dominant logic of the field. According to Bourdieu \cite{Bourdieu_1986}, capital has four forms: economic (e.g., financial resources), cultural (non-financial assets, such as physical appearance, spoken language, academic achievement), social (e.g., an individual's social network), and symbolic (prestige and recognition, such as awards). Individuals who begin their life with more capital, be that through inheritance or immediate exposure to the dominant culture, will be more able to gain personal and social advantages. For example, a student who is born into a family that speaks the dominant language of an educational institution may find it easier to learn, and a student with greater economic wealth may be more able to afford the costs associated with tertiary study (e.g., tuition fees, relocation, travel). The value of capital is not solely determined by form, but also by factors such as the manner of acquisition, and the personal characteristics of the owner. Issues emerge when an individual's capital is devalued unjustly by the `rules' operating in the field. For example, international research has shown that female physicists tend to receive fewer opportunities and career enhancing resources compared to objectively equal male physicists \cite{Ivie_2013}. Previous research of tertiary students suggests that female students may be more likely to discontinue physics education, regardless of performance \cite{Ellis_2016,Ost_2010}. Since disparities in enrollment still exist even after controlling for academic achievement, it is likely that capital is not the dominant factor in driving gender differences in physics education outcomes. Research does suggest, however, that the gender disparities in physics enrollments can be understood in terms of students' identity \cite{Brown2010,Hazari_2010,Hazari2013,Brewe_2016,Hazari2017} and self-confidence \cite{Litzler2014,Concannon2009,Sharma_2011,Beth_Kurtz_Costes_2008,Sawtelle_2012} -- factors that can be tied to Bourdieu's concept of habitus.

\subsection{Habitus}
Capital, in its various forms, interacts with habitus (Fig~\ref{fig:Fig1}); a construct defined by Bourdieu as a ``system of dispositions''\cite[p.170]{Bourdieu1984} formed in relation to a field. Whilst capital is what determines one's position within the field, habitus is what determines one's disposition towards it \cite{Bourdieu1992}. An individual's habitus is the internalization of the socio-cultural and historical context of a field, and it operates ``below the level of consciousness and language'' \cite[p.466]{Bourdieu1984}. Roy Nash understood habitus as ``a system of schemes of perception and discrimination embodied as dispositions reflecting the entire history of the group and acquired through the formative experiences of childhood''\cite[p.177]{Nash1999}. In simple terms, habitus is what we use to determine whether the field is something we are interested in, based on evidence present in the environment. Whilst habitus is generally formed during childhood within the family \cite{Dimaggio1982}, it is continually reconstructed and transformed as an individual operates in society. For example, a student who grows up in a family that places high value on science may share the same disposition \cite{archer2013aspires}. However, an individual may not choose to pursue science when faced with evidence that the field is not for them (for example, receiving poor grades, being treated poorly, lack of role models). Based on this internal matrix of dispositions, an individual's lifestyle practices are generated. According to Bourdieu, the collection of each individual lifestyle produced by habitus then constitutes the ``represented social world''\cite[p.170]{Bourdieu1984} --- the way that things appear to be. As the representation of the social world also influences the formation of habitus, the world and habitus share a reciprocal relationship. This relationship facilitates the cultural reproduction of inequity over time. 

Habitus can be used as a concept to explain the gender disparities in science enrollments. Based on what they see in their represented social world, students will ``[refuse] what they are refused (`that's not for the likes of us'), [adjust] their expectations to their chances, [and define] themselves as the established order defines them.'' \cite[p.471]{Bourdieu1984}. Based on what students see in their environment, they will make decisions on what they feel is a realistic study choice. Archer and colleagues explain this idea further: ``social axes of `race'/ethnicity, social class, and gender all contribute to shaping what an individual perceives to be possible and desirable.''\cite[p.885]{Archer_2012} The manner by which students perceive the different scientific disciplines, as they are represented in society, likely plays an important role in influencing their desire to study those disciplines. 

A wealth of research has outlined the various ways that women are subjugated in certain STEM disciplines (especially physics), with the culmination of these factors being referred to as ``the smog of bias'' \cite[p.1]{Kost_Smith_2010} or the ``gender filter'' \cite[p.370]{Blickenstaff_2005}. No single factor can sufficiently explain why women are less likely to pursue physics \cite{Kost_Smith_2010}, but a host of factors are likely to interact and impact on the dispositions students hold (habitus). Due to the pervasiveness of these various factors across society, habitus can take on a collective quality where individuals tend to hold stereotypical views on what is expected for members of different groups. To provide a simple example, research across 34 countries has shown that science tends to be implicitly associated with men more than with women, and that this level of gender bias predicts gender differences in science performance \cite{Nosek_2009}. As outlined by Bourdieu, an objective class of individuals can be considered the ``the set of agents who are placed in homogenous conditions of existence imposing homogenous conditionings and producing homogenous systems of dispositions capable of generating similar practices'' \cite[p.101]{Bourdieu1984}. In more basic terms, individuals who share similar backgrounds and characteristics will have a similar habitus, and this may predispose them to behave in similar ways \cite[p.434]{Reay_2004}. 

Every student holds beliefs about their possible educational paths. However, these beliefs are informed, implicitly and explicitly, by evidence in the environment. When deciding on whether to pursue physics, a student may ask: how are people like me treated in physics? Do people see me as a physicist? How many people like me study physics? Whilst we acknowledge that this is not an exhaustive list of reasons why students study physics, the answers to these questions are likely skewed to favour male students over female students. A study by Ong \cite{ong2005body} highlighted the incongruence felt by minority female physics students as they studied physics, where their competence was unfairly questioned because their `bodies did not fit' with the stereotypical depiction of the white male scientist. Similarly, studies have found that women are more likely to be viewed as incompetent, controlling for confounding variables other than gender, by scientists (including physicists) looking to hire a laboratory manager \cite{Moss_2012}, or by students evaluating their physics teacher \cite{Potvin_2016}. Similarly, women rated as more feminine are less likely to be judged as a scientist \cite{Banchefsky2016}. The pervasive nature of the ``smog of bias'' \cite[p.1]{Kost_Smith_2010} in physics offers the `homogenous conditions of existence' that may result in a gendered habitus in physics: one that sees physics as unwelcoming for female students. This is likely to explain why studies tend to find that female students are more interested in life science subjects \cite{Buccheri_2011} and male students are more likely to be interested in physics, engineering and mathematics \cite{Su_2009, B_E_2013, Cunningham_2015}. It is important to note here that the opposite is not true --- there is a lack of evidence to suggest that male students are unfairly judged as incompetent or feel unwelcome in the life sciences and therefore choose physical science subjects.  

Evidence suggests that the gender differences in subject interest may not be present in early childhood, but emerge by the end of secondary school \cite{Baram_Tsabari_2010}. This lends credence to the idea that habitus is formulated over time; as individuals become increasingly aware of societal norms, their interests align with what (through their habitus) seems like a realistic study choice. These stereotypical gender preferences may persist when it comes to the types of science-related career that secondary school students aim for \cite{Kj_rnsli_2011}, and students' choice of STEM major at university \cite{Bottia_2015, Sadler_2012}. At university level, gender disparities may even widen further; a study of physics students at a university in the United States found that female students are more likely to see their interest in physics diminish during introductory physics \cite{Kost_Smith_2010}.

\section{The Current Study}

The current study was motivated by the need to understand any potential gender differences in student outcomes in general, and at the University of Auckland (UoA) in particular. Our study seeks to not only understand the outcomes for physics students at the UoA, but to employ a unique approach that highlights the complexity of student enrollments and places them in a wider socio-cultural context. To do so, we employ network analysis on student enrollment records to provide a detailed representation of the field of physics at the UoA. The network analysis approach builds on the work of \cite{de2003fields} and \cite{bottero2011worlds} who described the utility of combining network analysis with Bourdieu. \cite{bottero2011worlds} provide an example of how networks of social relations can provide a representation of a field. The current study expands on this area of research by conceptualising academic fields as communities detected in networks of course selection. Furthermore, we draw attention to under-utilised concepts of Bourdieusian theory: the concepts of transverse movement between fields, and vertical movements within fields. We focus on providing a basic description of the movements that physics students make within and between academic fields at the UoA. Our study echoes previous studies that analyse the pathways that students take through education. However, by combining the network analysis approach with the sociological theory outlined by Bourdieu\cite{Bourdieu1984}, w meove beyond simple models to a more nuanced description of the way habitus can be depicted/demonstrated through network analysis as both a cause and a symptom of gender stratification. 

\subsection{Transverse and Vertical Movements}
Bourdieu's theory encourages us to view student movements across STEM domains in relation to the structures of the field, the volume of capital a student holds, and the manner by which habitus guides practices in the field. In addition to our objective representation of the field of physics, we also consider what may motivate these movements, based on evidence from previous research.

According to Bourdieu, society is structured in a manner that allows individuals to engage in two types of movement: vertical and transverse: ``vertical movements, upwards or downwards in the same vertical sector, that is in the same field... and transverse movements, from one field to another, which may occur either horizontally or between different levels'' \cite[p.131]{Bourdieu1984}. Vertical movements upwards require an increase in the prized capital in the field. In tertiary science education, this may be represented by grades in science courses over time. Transverse movements entail a shift to a new field, and the conversion of accumulated capital into the capital accepted in the new field. For example, a student making a transverse movement from physics to life sciences will have to assimilate to a different skill set, and even a different culture. Transverse movements can be used as a strategy to protect a relative vertical position: 
\begin{quote}
    ``transverse movements entail a shift into another field and the reconversion of one type of capital into another or of one subtype into another subtype... and therefore a transformation of the asset structure which protects overall capital volume and maintains position in the vertical dimension''\cite[p.132]{Bourdieu1984}
\end{quote} When an individual feels that they are slipping in the ranks of the field, they may choose to make a transverse movement to a new field, where their accumulated capital holds more translatable value. 

In the current study, we conceptualize cultural capital in its institutionalized form as measured by course grades. The current study, therefore, seeks to understand:
\begin{itemize}
    \item Whether there are gender differences in UoA physics students moving from one academic field to another.
    \item Whether there are gender differences in the persistence of UoA students in physics.
    \item Whether there are gender differences in UoA physics students moving upwards or downwards in academic achievement (as signalled by course grades).
\end{itemize}

Whilst our data do not allow us to conceptualize forms of capital other than institutionalised cultural capital (i.e, course grades), our methodology leaves the opportunity for future research to incorporate other measures of students' capital. More specifically, future research should investigate how other forms of capital are distributed across fields and relate to the movements that students make. 

\section*{Materials and methods}

\subsection{Data}
The current study uses administrative student data from the UoA from 2009 to 2014 (N $=$ 8905), including demographic and academic information. For the purposes of this study, the only demographic variable considered in the analysis was gender. Academic variables include course codes that students were enrolled in, and the year and semester in which they were enrolled. We did not have information regarding students' degree plans or majors. Records of non-physics courses were included as long as a student had enrolled in at least one physics course during the study period. At the UoA, students are required to take two courses outside of their major, with the options being titled as general education courses. We excluded all students who studied a general education course in physics from our analysis. We know that these students are not physics students, and they do not offer a representative sample of students from outside of physics.

A typical Bachelor of Science physics degree at the UoA takes place over the course of three years. In their first year, physics students are required to take Advancing Physics 1 (AP1) and then Advancing Physics 2 (AP2) before moving onto second year physics. Life science students (those majoring in biomedical sciences or medicine) are required to take Physics for Life Sciences (PLS) in their first year. PLS is taught by the physics department. This means that, despite our study population including only students who took a physics course, many of the students present in our data set were likely majoring in life sciences. Our population therefore allows us to compare the outcomes for students in the physics and life sciences disciplines. AP1 and PLS cover the same content, but are presented in a different manner. One significant difference between AP1 and PLS is that AP1 assumes a knowledge of calculus, while PLS does not. This is an important point to consider, as a mathematics background may be an important form of science related capital \cite{Black2016}, and female students may be more likely to drop out of physics education after taking calculus \cite{Ellis_2016}. The current study was able to compare the AP1 and PLS subsets of the general physics population to account for a student's first year disciplinary intentions. PLS is still considered an acceptable prerequisite for AP2 in lieu of AP1, although it is rare for students to take this route.

\subsection{Measures}

The following variables were used in the analysis:
\begin{itemize}
\item{Grade Point Equivalence (GPE):} GPE is an entry level score that provides a standard measure of a student’s prior academic performance at the time of admission to university, regardless of the qualification they previously took. It is measured on a 0-9 scale, with 9 being the highest performing. It provides an aggregate measure of how well a student did in all of their high school courses \cite{UoA_2016}.

\item{Grade Point Unit (GPU):} GPU is a measure of a student’s university performance in a single course. It is measured on a 0-9 scale, with 0 being equivalent to a fail (D+ or lower), and 9 being equivalent to an A+ grade. GPU was used as a measure of performance for AP1, AP2 and PLS.
	
\item{Gender:} Due to limitations in the administrative data that were used, gender was only recorded as male or female. 

\end{itemize}

\subsection{Procedure}
Although Bourdieu offers a rich theory to interpret movements within and between fields, we are left with the challenge of defining what constitutes a field. Whilst it could be argued that every student who takes a physics course at university is a physics student, we believe that this is not sufficient. Students may be enrolled in a subject discipline on paper, but actually be fully engaged in a separate field of study. A good example of this is PLS. PLS students may be considered physics students on paper, but their main field of study is likely biomedical sciences or medicine. Through network analysis, we are able to define academic fields in terms of the patterns of course selection. We represent course selection patterns as a network, where nodes represent university courses and edges represent the enrolments of students within courses. We then explore the structure of the network by investigated the communities of courses that tend to be taken together by students. Our approach, similar to blockmodelling approaches \cite{bottero2011worlds,white1976social}, allows us to take a complex network and reduce it to its core structure. It does this by identifying communities of nodes that tend to share more edges. We can then explore patterns at the level of communities instead of at the level of nodes. In the current study, we interpret these communities as academic fields. Following this, we are able to investigate gender differences in the transverse that students make across the fields represented in our network. We supplement our network with course achievement data to compare vertical movements within and across fields. 

The following section outlines the series of steps that were used to generate the course network and use it to answer our research questions regarding gender differences in students transverse and vertical movements. Through the analysis of course relationships, we can take a non-biased approach to defining the fields in which students are located. 

\subsection{Forming the Network}
To begin our network analysis, we structured our data as an adjacency matrix, where both rows and columns represent the courses taken by students in our sample, and a cell value is the number of students who took both course $i$ and course $j$ within their undergraduate degree. Whilst we could define edges in relation to the frequency of students who took a pair of courses, this does not truly reveal the underlying community structures we are interested in. Pairs of courses including one very popular course will tend to have higher values regardless of whether the two courses belong to the same academic field. We take into account course populations by normalising the matrix using a Revealed Comparative Preference (RCP) score. RCP measures the fraction of students from a course $j$ who also took a second course $i$, relative to the overall fraction of students taking course $i$, across all other courses. More specifically: 
$$RCP(i,j) = \frac{x_{ij}/x_j}{x_i/x}$$
where $x_{ij}$ is the number of students taking both course $i$ and $j$, $x_j$ (or $x_i$) is the total number of students taking course $j$ (respectively, course $i$), and $x$ is the total number of unique students enrolled in any course.
The RCP metric is based on the measure Revealed Comparative Advantage, used in economics \cite{Balassa1965}, and was calculated using the EconGeog package in R \cite{balland2017economic}. The RCP approach to normalising gives the ``revealed'' course preferences, controlling for the enrollment numbers of each course (that is, the courses that tend to be taken together by students in the network more often than would be predicted by the course populations alone). RCP values greater than one indicate that a pair of courses had a `preference' for being taken together, given the relative populations of both courses, whilst RCP values below one indicate no evidence of any preference. Thus, we exclude any network edge with an RCP weighting lower than one, leaving only the course pairs that had a preference for taking together. 

We identify communities of courses that tended to be taken together by students. We employ the community detection algorithm Infomap \cite{Edler_2017}) on the network. In basic terms, this method of community detection reveals communities of nodes based on maximising a modularity score. In network analysis, modularity is the extent to which a network is partitioned so that the number of edges within communities is greater than the number of edges between communities. Using the igraph package in R \cite{csardi2006igraph}, the infomap algorithm identified 23 communities of courses in our network (see Table \ref{Table 1}). Each community can be interpretted as a unique academic field consisting of different combinations of courses and requiring different sets of knowledge. The resulting network is shown in Fig \ref{fig:Fig2}. Nodes in the network represent courses taken by students, whilst edges show a preference for a pair of courses being taken together. Node colours represent the communities of courses, which we interpret as individual fields. Using Bourdieu's concepts of transverse and vertical movements, we explore the relationships between and within the 23 communities (or fields) in the network.

\begin{table}[!ht]
\caption{\label{Table 1} \textbf{Compositions of the Communities Detected in the Course Network.}}
\begin{tabular}{|c|c|c|c|}
\hline
Community & Count students & Proportion Female & Total Enrollments \\ \hline
Ancient History & 150 & 0.48 & 205 \\
Biological Science & 5630 & 0.52 & 18600 \\
Chemical Materials & 20 & 0.35 & 60 \\
Chemistry & 1660 & 0.49 & 4180 \\
Chinese & 60 & 0.27 & 85 \\
Computer Science & 4405 & 0.28 & 22195 \\
Engineering & 980 & 0.36 & 9315 \\
Finance-Marketing & 1430 & 0.29 & 6735 \\
Food Science & 640 & 0.53 & 1505 \\
Geography-Geology & 1470 & 0.38 & 6095 \\
Japanese & 85 & 0.38 & 180 \\
Law & 170 & 0.46 & 240 \\
Liberal Arts & 6410 & 0.38 & 12750 \\
Medical Science & 4715 & 0.53 & 26790 \\
Nursing & 70 & 0.81 & 300 \\
Optometry & 550 & 0.61 & 1310 \\
Pharmacy & 645 & 0.58 & 2100 \\
Physics-Maths & 3060 & 0.25 & 12125 \\
Population Health & 200 & 0.57 & 510 \\
Psychology & 1440 & 0.52 & 4410 \\
Sports Science & 245 & 0.47 & 575 \\
Statistics & 1470 & 0.36 & 3985 \\
Surgery & 350 & 0.43 & 2800 \\ \hline
\end{tabular}
\textit{Counts have been rounded to the nearest 5 to preserve confidentiality. Proportions were formulated using original values.}
\end{table}

\subsubsection{Transverse Movements}
To investigate whether there are gender differences in UoA physics students moving from one academic field to another during their undergraduate degree (transverse movements), we build on the network outlined in the previous section. We take the same set of nodes, with the same community structures, but weight edges by the number of students who took course $i$ \textbf{before} course $j$. From this new directed network, we are able to assess the movements that students make between communities. To answer our questions regarding the transverse movements that students make from one field to another, we aggregate the number of movements from courses within community $m$ to courses within community$n$ (see Fig~\ref{fig:Fig3}). For example, the courses in the \textit{Physics-Maths} community in Fig~\ref{fig:Fig2} become a single Physics-Maths node in Fig~\ref{fig:Fig3}. Outgoing edges between communities are aggregated into a single outgoing edge, with a weighting equivalent to the sum of all outgoing edges weights from nodes in the community. Edges between courses within a community are similarly aggregated, and are represented as self-loops (a link from a node to itself) in Fig~\ref{fig:Fig3}. To investigate how transverse movements differ by gender, we calculate the odds (with 99\% confidence intervals) of a female student moving from community$_i$ to community$_j$ over a male student. The new network is represented as a network in Fig~\ref{fig:Fig3} and as a heat map in Fig~\ref{fig:Fig4}. 

\subsubsection{Vertical Movements}
We also seek to investigate how male and female students with differing levels of prior achievement choose to invest their capital. Are there gender differences in the vertical movements (moving upwards or downwards in the objective rankings in a field) that students make from one stage to the next? Do male and female students with different levels of prior achievement choose to invest their capital differently? To understand the nature of students' vertical movements between within and between fields, we incorporate student achievement data into our previously established network. For each course, we have the student grade point unit score (i.e., their level of achievement). Our data set also includes an average high school achievement measure, called Grade Point Equivalence (GPE), for the majority of students in our network. This allows us to look at the transitions that male and female students make from high school to university study. 

We are particularly interested in the movements that students make going from high school to three specific stage one courses: AP1, AP2, and PLS. We also investigate the gender differences in vertical movements that students make from these physics courses to our detected fields, and between our detected fields. For our detected fields we calculate a Grade Point Average (GPA) score for each student, in which we take the mean of the student's grade point unit scores for each course they took within the community. For example, the Physics-Maths GPA score will be a student's mean average grade point unit score for all of the courses they took within the Physics-Maths community. 

As outlined by Bourdieu, an individual's power in a field is determined by the composition and volume of capital they hold \textit{relative} to other individuals. As our goal is to compare the relative vertical position of students within and between fields, we convert the achievement scores (GPE for high school, GPU for the key stage one courses, and GPA for the communities) into percentile ranks. Standardising achievement in this manner facilitates comparisons across fields. Top achievers in a field will have a percentile rank score of 100, whilst low achievers will have a percentile rank score closer to 0. We can then compare the change in percentile rank scores for male and female students across our network. To describe the gender differences in vertical movements, we use independent 2-group Mann-Whitney U Tests (a non-parametric t-test) to compare differences in percentile rank change between male and female students. We were then able to determine whether there were any significant differences between male and female students gaining in relative performance across fields. We also report the odds (with 99\% confidence intervals) of top, middle, and low achieving female students enrolling in different fields (where achievement groups are based on percentile rank split into three equally sized bins). We explore the movements from high school to key stage one university courses specifically, and from key stage one physics courses to detected fields. 

\section{Results and Discussion}
Networks showing the revealed communities of courses that students take can be seen in Fig~\ref{fig:Fig2} and Fig~\ref{fig:Fig3}. Fig~\ref{fig:Fig2} shows the network of courses offered by the University of Auckland (UoA) between the years 2009 and 2014, with communities indicating courses that tend to be taken together within students' undergraduate degrees (represented by the different colours). The communities include (in ascending order of aggregated course enrollments): Medical Science, Computer Science, Biological Science, Liberal Arts, Physics-Maths, Engineering, Finance-Marketing, Geography-Geology, Psychology, Chemistry, Statistics, Surgery, Pharmacy, Food Science, Optometry, Sports Science, Population Health, Nursing, Law, Ancient History, Japanese, Chinese, and Chemical Materials.

The use of Revealed Comparative Preference (RCP) in conjunction with the community detection reveals underlying academic fields in which physics student participate, as indicated by the combinations of courses that students enroll in. Physics courses (including AP1 and AP2, the first prerequisites for a physics major at the UoA) and mathematics courses are located the same field, which we label \textit{Physics-Maths}. PLS, a physics course required for students wanting to study medicine, belongs to the field of \textit{Medical Sciences}. We report the the counts of students per community, with the percentage of female students, in Table \ref{Table 1}. Liberal Arts, Biological Science, and Medical Science were the three largest communities based on number of unique students enrolled in the field. Medical Science, Computer Science, and Biological Science were the largest communities in terms of total enrollments (an individual student may be enrolled in more than one course per field). In terms of the proportion of female students per community, Physics-Maths (0.25), Computer Science (0.28), and Chinese (0.27) were the most male dominated. Nursing (0.81), Optometry (0.61), and Pharmacy (0.58) were the most female dominated. 

The network and RCP approach provides a non-biased method of classifying the fields in which students are participating in. The use of RCP shows that disciplinary labels (i.e., `Physics') are imperfect in classifying the patterns of courses that students enrol in. Although PLS is a physics course, it has a higher affinity with the life sciences, and our community detection approach reflects this by locating PLS within the field of Medical Science. For example, the percentage of female students enrolled in \textit{all} physics courses (including PLS) is 40\%. Our community detection shows that female students only make up around 25\% of the main Physics-Maths community. The difference between these percentages is substantial, and raises important implications for the way in which universities report the number of students studying in different disciplines.

\subsection{Transverse Movements}
We first wanted to understand whether there are gender differences in UoA physics students moving from one academic field to another. Our results regarding these transverse movements (more detail is given in the supplementary material) show that female students were around 1.82 (CI: 1.63-2.02) times more likely to take a course in Biological Science after taking a course in Physics-Maths, and 1.44 (CI: 1.40-1.47) times more likely to take a further course in Biological Science after taking a previous Biological Science course. On the other hand, male students were around 1.97 (OR $=$ 0.51, CI: 0.48-0.56) times more likely to take a course in Physics-Maths after taking a course in Biological Science. There were no significant gender differences in students taking a course in Physics-Maths after taking a previous course in that community. Male students were consistently more likely to take a course in Computer Science after taking a previous course in another community, for example going from Biological Science (OR = 0.44,CI: 0.42-0.46), Physics-Maths (OR $=$ 0.52 , CI: 0.50-0.54), and Computer Science (OR $=$ 0.44, CI: 0.43-0.45).
 
The above results show differences in the transverse movements that students make between fields. Female students were nearly twice as likely to switch into Biological Sciences after Physics-Maths, with male students nearly twice as likely to go the opposite direction. Thus, the results of the current study show that gender disparities are evident, not only in the fields in which students choose to study (see Table~\ref{Table 1}), but also in the transverse movements make between fields. These findings are in line with previous research that shows that the life science disciplines (biology, medicine etc.) tend to be more popular for female students, while physics, maths, computer science and engineering tend to be more popular for male students \cite{NSF, Cunningham_2015,Kost_Smith_2010,Heilbronner_2012, Stevanovic_2013, InstituteofPhysics_2012, Smith_2011, EducationCounts_2016a, Kennedy_2014, Semela_2010}. Male students were consistently more likely to switch into computer science regardless of prior field. This highlights the field of computer science as a key area of future investigation.

Whilst the above findings indicate gender disparities in student enrollments, it is important to consider the achievement levels of students who enter in to different fields, and whether achievement impacts on the movements that students make. We want to know whether there are gender differences in the persistence of UoA students in physics, accounting for student achievement. The question we now ask is where do male and female students with differing levels of prior achievement choose to invest their capital? We look specifically at students coming from high school, to the key stage one physics courses (AP1, AP2, and PLS) and to the disciplinary fields revealed in our network. 

\subsection{Vertical Movements}
We investigate the impact of student achievement on student enrollment in two ways. We firstly report the number of top, middle, and low achievers who make movements from high school to the key stage one physics courses (AP1, AP2 and PLS), and to the fields revealed in our network. We then assess the vertical movements that students make by analysing gender differences in the change in objective rankings within and between fields. We begin by reporting the progression of students from high school to university physics.

For students ranking in the bottom third of high school students (``low achievers''), 17.94\% of female students and 35.53\% of male students went on to study AP1. Thus, low achieving male students were 2.52 (OR $=$ 0.40, CI :0.30-0.52) times more likely to enter the main physics pathway at the UoA compared to their female counterparts. For students ranking in the middle third of high school students (``middle achievers''),  10.69\% of female students and 30.58\% of male students went on to study AP1. Thus, middle achieving male students were 3.68 (OR $=$ 0.27, CI: 0.20-0.37) times more likely to enter physics at the UoA compared to their female counter parts. 

Our findings show that of the students who were ranking in the top third of students coming from high school (``top achievers''), very few chose to invest their capital in physics. Only 8.72\% of male students and 5.06\% of female students who were top achievers from high school chose to enrol in AP1. These percentages also indicate that from this top achieving group, male students were 1.79 times more likely to go to AP1 (OR $=$ 0.56, CI: 0.36-0.87). Thus, not only does it appear that physics is an unattractive option of top achieving high school students, but this is particularly true for top achieving female students. In contrast, 72.66\% of male students, and 88.35\% of female students from this top achieving group enrolled in PLS. Top achieving female students were thus 2.84 (CI: 2.10-3.83) times more likely than their male counter parts to follow this pathway. 

Top achieving students are most able to make choices on where they choose to invest their capital, and thus our results provide a good indication that the life science fields tend to be viewed as high in symbolic capital (prestige). The fact that we did find differences in the choices to study PLS over AP1 suggests that physics is viewed as a less rewarding study path than the life sciences. Questions need to be asked about the way in which physics is presented to students in secondary school. Claussen and colleagues argue that science education needs to highlight the utility value of science in culture, scientific literacy, and employment \cite{Claussen_2013}. Students will choose to invest their capital in a field where they feel that they can get the largest return (be it in educational qualification, future employment opportunities, or enjoyment). Our findings suggest that within science education, physics needs to make a stronger case for its utility in order to attract high achieving students, in general, and female students in particular. This could be achieved by boosting science capital\cite{Archer2015a}, increasing the knowledge about the future value of physics courses in the employment market \cite{Archer2014}, and providing information on the utility of physics in everyday life. 

Increasing the value of physics and boosting the related capital of students within physics, although necessary, is likely an insufficient strategy to address gender disparities. In the context of previous research, we also argue that female students' choice is constrained due to the unwelcoming climate presented in the field of physics \cite{Blickenstaff_2005, Kost_Smith_2010}. Following Bourdieu's theoretical framework (Fig \ref{fig:Fig1}, we must also consider students' habitus. The affinities that students feel towards each scientific discipline is influenced from an early age by their experiences in, and perceptions of the field of science education. Using evidence from their life experiences, students enter into university with an idea of what discipline is `\textit{for me}'. For fields such as physics and computer science (where we found the most consistent gender disparities in enrollments) students are likely influenced from an early age by the ``smog of bias'' \cite[p.1]{Kost_Smith_2010} that targets women. Through the combination of a myriad of factors, from the negative gender stereotypes \cite{Nosek_2009}, to the ways in which women's competence is unfairly questioned \cite{Moss_2012,Potvin_2016, ong2005body}, students will internalise (via habitus) the perception that physics is something men do, and where women are unwelcome \cite{Archer_2013}. Until the `smog of bias' is addressed, female students will continue to have constrained choice in science.

Whilst we could interpret the lower likelihood of a male student studying in the life sciences as evidence of constrained choice also, we find this an unrealistic interpretation. The sizable representation of male students and researchers in the life sciences presently and historically, and the lack of negative factors that impact male students in this domain, mean that the life sciences are likely still a realistic study choice for male students. To put more simply, male students have more choice on where to invest their capital, whilst female students are more likely to face obstacles. The rules operating in the field of physics may require female students to make extra effort to appear competent and persevere in the field. As outlined by Ong \cite[p.594]{ong2005body} in a study of minority female physics students: ``the ways in which women of color organize themselves to appear competent in the context of physics specify invisible rules about the strict boundaries around local scientific communities.'' The idea that women in physics may have to ``relegate social and cultural identities to the margins'' \cite[p.597]{ong2005body} in order to succeed in physics corresponds to Bourdieu's idea that individuals lacking in the `valued' cultural capital in a field may need to make sacrifices to get ahead \cite[p.333]{Bourdieu1984}. 

Of the students from AP1 who ranked in the bottom third of achievers (low achievers), we found that 40.38\% of male students, and 28.29\% of female students progressed to AP2. Thus, female students from this low achieving group were around 1.717 (OR $=$0.58, CI: 0.35-0.98) times less likely to progress from AP1 to AP2. There were no significant gender differences in the middle (OR $=$ 0.84, CI:0.56-1.24) and top achieving (OR $=$ 0.77, CI: 0.49-1.21) AP1 students who went to AP2.

The above findings point to previous research that suggests that female students may be less confident in physics \cite{B_E_2013,Sharma_2011,Hofer_2016} and maths \cite{Else_Quest_2013, Sheldrake_2015}, or, rather, low achieving male students may be \textit{over-confident}. It may be that in our sample, gender differences in progression from AP1 to AP2 for middle and top achieving students were not present as the grades received offered evidence that they \textit{belong} in physics. For the low achieving students, belonging is not evidenced by their grades. Low achieving male students may be buffered by a habitus that, after years of socialization, predisposes them to physics. Female students, on the other hand, may be less likely to have this protective disposition. Whilst further research is needed to substantiate this claim, past research does suggest that students are more likely to make internal attributions of failure for female students in science (i.e., they fail because they are not good at it), and external attributions of failure for male students (i.e., unfavourable circumstances) \cite{LaCosse_2016}. Furthermore research by Ellis, Fosdick and Rasmussan\cite{Ellis_2016} found that female students are more likely to discontinue physics after taking an introductory calculus course, with female students also being more likely to cite lack of understanding as a reason for dropping out. This may also apply to students in our sample, as AP1 includes content that requires knowledge of calculus.

We also investigated the rank change for students moving from high school to the key stage one physics courses, and from those physics courses on to the fields detected in our network. We found no statistically or practically significant gender differences in the vertical movements in these pathways, with the exception of students going from high school to AP1. As indicated by Fig~\ref{fig:Fig5} and \ref{fig:Fig6}, we found that low and middle achieving female high school students were more likely to decrease their rank in the field (i.e., make a vertical movement downwards) in AP1 with this being significant. On average, low achieving female students went down around 6 ranks compared to their male counterparts (Difference in Position=$-5.71$, CI: $-8.57$ - $-2.86$), while middle achieving female students went down around 9 ranks relative to their male counterparts (Difference in Position=$-8.57$, CI: $-12.86$ - $-2.86$). There was no significant gender difference in rank change for top achievers. 

These results somewhat echo the findings of Kost-Smith \cite{Kost_Smith_2010}. They found that male students tended to outperform female students on post-test physics concept inventory scores, despite there being no gender differences in pre-test scores. Based on this, we would expect male students in our sample to also increase their relative position in the field of physics after first year study. With that being said, the gender differences we did find were relatively small, and non-existent for top achieving students. The top achieving female students in our sample who chose to progress in physics likely have a habitus that is just as congruent with physics as the male students (i.e., they feel that physics is `for them'). However, taken in the context with our other findings that female students were less likely to progress from high school to AP1, or from AP1 to AP2, questions must be asked about the distribution of physics-related capital and the development of physics habitus \textit{before} university education, particularly for middle and low achieving students. Many studies point to the late childhood and early teenage years as a key formative stages \cite{archer2013aspires,DeWitt2014,Baram_Tsabari_2010} for identity within science. Future studies of tertiary education in New Zealand should investigate the role of science identity in subject selection decisions further.

\subsection{Implications}
The current study offers a detailed account of the movements that students make through university physics. Our results show that female students were less likely to progress from high school to AP1, regardless of prior achievement, while low achieving female AP1 students were less likely to progress to AP2. The findings of the current study suggest that more needs to be done to ensure that physics is perceived as a viable option for female students and high achieving students (and particularly high achieving female students). This can be done by using interventions to boost the value that science capital holds in all areas of society. Echoing the arguments of Claussen and Osborne\cite{Claussen_2013} and Archer and colleagues \cite{Archer2012}, science education in New Zealand, and internationally, needs to highlight the utility value of physics in culture, in boosting scientific literacy, and employment. 

However, we argue that boosting the value and access to capital, despite being a necessary goal for boosting the numbers of students in physics, is insufficient to tackle gender disparities. Following our research framework, we should seek to transform the habitus of students to encourage them to invest their capital in physics. We need to continue to change the culture of physics so that it is more likely to be viewed as a viable study option. Whilst we do not want to force students to study in areas where they do not want to be, we echo the sentiments of Cheryan and colleagues, who state: `
\begin{quote}
Just because women are excited to go into other fields does not mean that they would not have been equally excited to go into computer science, engineering, and physics if the cultures signaled to them that they belong there.
\end{quote}\cite{cheryan2017some} We need to transform the field of physics so that it signals to female students that they belong there. 

Previous research suggests that interventions to boost the number of female students graduating in physics would be most useful at stages of education prior to university \cite{cheryan2017some}, as intentions to study science can be formed by early secondary school \cite{archer2013aspires,Baram_Tsabari_2010}. Female students' self-concept in physics may be improved through exposure to supportive family members \cite{kelly2016social} and high school teachers \cite{Hazari2017, kelly2016social}. 

The results of the current study do also indicate that more needs to be done to support the female students who have already chosen to study physics at university. This may take the form of increased academic support for low achieving students in particular. Universities can seek to provide group learning experiences in introductory physics \cite{Sawtelle_2012}, and more welcoming environments for female students in physics and computer science \cite{master2016computing}. 

As outlined by Bourdieu, individuals fight to define the criteria of what is of value in the field. Individuals who hold power in the field have the means to change the culture of physics. As stated by \cite[p.11]{hilgers2014introduction}:
\begin{quote}
    The chances that established actors will succeed in preserving the order [of the field] are, however, greater than the probability of subversion. The more legitimate an agent, the more her peers consume her products, and the more they consume her products, the more legitimate she becomes.
\end{quote}
Following this logic, culture change in the field may require forced institutional changes. Initiatives to help address the inequities faced by women already in the field \cite{Ivie_2013}, and to increase the representation of women in research and higher education \cite{AthenaSwan,Horizon2020} are important steps to fostering changes in culture. through these initiatives, we signal to future students from all backgrounds that physics is somewhere where women belong. 

Beyond our research findings, the current study demonstrates the utility of using network analysis and Bourdieu together. Whilst network analysis serves as a good method for representing Bourdieu's concept of field, Bourdieu's theory provides a rich interpretive lens. Our approach carries many benefits over other, more simplistic frameworks, such as the leaky pipeline. Employing the concepts of field, capital and habitus allow us to understand the objective structure of physics, whilst respecting the subjective contexts in which students are placed. Doing so removes stigma that can be attached to students who `leak' from the physics education pipeline. Emphasising the contexts that students are situated in allows us as researchers to place our findings in a broader context and formulate suitable interventions to boost the physics enrollments of underrepresented groups. 

\subsection{Limitations}
There are limitations to the current work that future studies should address. Firstly, our data set is limited to UoA physics students, and included only course selection and performance information, and minimal demographic information. We did not have data regarding the course selection information of students prior to university, whilst our measure of high school achievement was a general measure and not subject specific. More detailed data would have provided more information regarding students' educational trajectories. With that being said, our results show the utility of working with student record data. Our network analysis, whilst simple, also provides a strong framework for working with more complex data; for example, investigating the distribution of economic, cultural, social capital across the network.

Whilst we argue that our network analysis approach enables us to draw many conclusions from our data, our study would also have benefited from combining our quantitative analysis with qualitative measures. We have used a quantitative approach to defining the field, and used evidence from other research studies to draw conclusions from our data. Whilst this approach is informative, qualitative approaches can provide even more context specific details. Bourdieu highlighted the need to break the dichotomy between the aim of understanding the `objective reality' (the overall distributions of groups and relationships between them) and the aim of understanding ``not `reality', but agents' representations of it'' \cite[p.482]{Bourdieu_1986}. Surveys and interviews of students would provide contextual and fine-grained detail that would complement our quantitative network analysis. Qualitative analysis may also be a more appropriate way to investigate gender as a non-binary construct.  

Despite having access to information regarding the ethnicity of students, we decided not to present this information in the current analysis. This is due to the fact that preliminary analysis showed low cell sizes for ethnic groups other than New Zealand European and Asian students in physics, in particular M\={a}ori and Pacific Island students (these findings are available on request). When possible, future studies should make use of an intersectional research design (one that explores the interaction between gender, ethnicity, social class etc.). This is especially important when using a Bourdieusian framework to interpret results. As suggested by Bourdieu: ``The individuals grouped in a class that is constructed in a particular respect... always bring with them secondary properties'' \cite[p.102]{Bourdieu1984}. Understanding the intersection of student characteristics would allow us to control for the secondary properties that Bourdieu speaks of. The authors are currently conducting further to understand why there were low cell sizes for minority groups using data from earlier educational stages (i.e., secondary school). 

\section*{Conclusion}
The current study investigated gender differences for undergraduate physics students at the University of Auckland (UoA) through the use of network analysis on student data, with an interpretive lens based on the work of Pierre Bourdieu. Our network analysis revealed the different academic fields in which students are situated. We outline the utility of networks in visualising Bourdieu's concepts of vertical and transverse movements within and across fields. Analysis showed gender differences in transverse movements (moving from one field to another) consistent with gender stereotypes: female students were more likely to enroll in life science fields (Biological Science, Medical Science), while male students were more likely to enroll in the Physics-Maths and Computer Science fields.Analysis of a UoA student-course network revealed that female high school students are more likely to study life sciences at university compared to physics, and this is particularly true for high achieving students in this group. Furthermore, of the female students who did enter physics in their first year, low achieving students in this group were less likely to progress to further physics compared to their male counterparts. We relate these findings to Bourdieu's concepts of field, capital and habitus (Fig~\ref{fig:Fig1}). High achieving secondary school students (especially female students) may see more of a return for their capital in the life sciences compared to physics. Whilst it may be that physics does a poor job of highlighting its value, we argue that female students will continue to suffer constraints in their subject selection until the `smog of bias'\cite{Kost_Smith_2010} in physics is addressed. As outlined by Kost-Smith and colleagues\cite{Kost_Smith_2010}, it is unlikely that a single factor can account for the gender disparities seen in physics enrollment. We suggest that the various factors that have been linked to the attrition of women from physics (e.g., negative gender stereotypes, lack of female role models etc.) culminates into a gendered habitus that increases the likelihood of students viewing physics as a field that men do and where women are unwelcome. We close by discussing potential avenues for addressing gender disparities, which focus on not only boosting access to, and the value of, physics related capital, but also transforming the culture of the field so that all students (regardless of gender) view physics as a feasible study option.

\begin{figure}[!ht]
\begin{center}
\begin{tikzpicture}[scale=1]

\node at (1,-0.0) {\textbf{Capital}};
\node at (2.1,-0.0) {\textbf{$\times$}};

\node at (3.3, -0.0) {\textbf{\textbf{Habitus}}};
\draw[very thick, ->] (4.5,-0.0) -- (7.6,-0.0);
\node at (6.0, 2.5) {\textbf{\textbf{Field of Study}}};

\node at (10,0.5) {\textbf{Practices/Dispositions}};
\node at (10,-0.0) {Grades};
\node at (10,-0.5) {Course enrollment};

\draw[black, very thick] (6,0) ellipse (7 and 3);
\end{tikzpicture}
\end{center}
\caption{\textbf{Simplified Bourdieusian Theoretical Model.\newline}
The Bourdieusian framework used in the current study, adapted from the original model outlined by Bourdieu \cite[p.10]{Bourdieu1984} and the work of Archer and colleagues \cite{Archer2012,Archer2014, Archer_2015}. A student's habitus interacts with their acquired level of capital (in particular science-related capital) to generate a student's practices (behaviours, grades etc.) and their dispositions towards the field. A student's habitus, a matrix of internal dispositions \cite{Reay2004}, is formed in relation to the specific socio-cultural and historical context of a field. A student who is positively predisposed to study in a scientific field, whilst also having access to various forms of science-related capital, will likely achieve higher grades in that field and aspire to study in that field in the future. A student who encounters bad experiences in the field will likely be dissuaded from future study via their habitus (`this discipline is not for me').}
\label{fig:Fig1} 
\end{figure}

\begin{figure*}[!ht]
\centering
\includegraphics[width = 13cm]{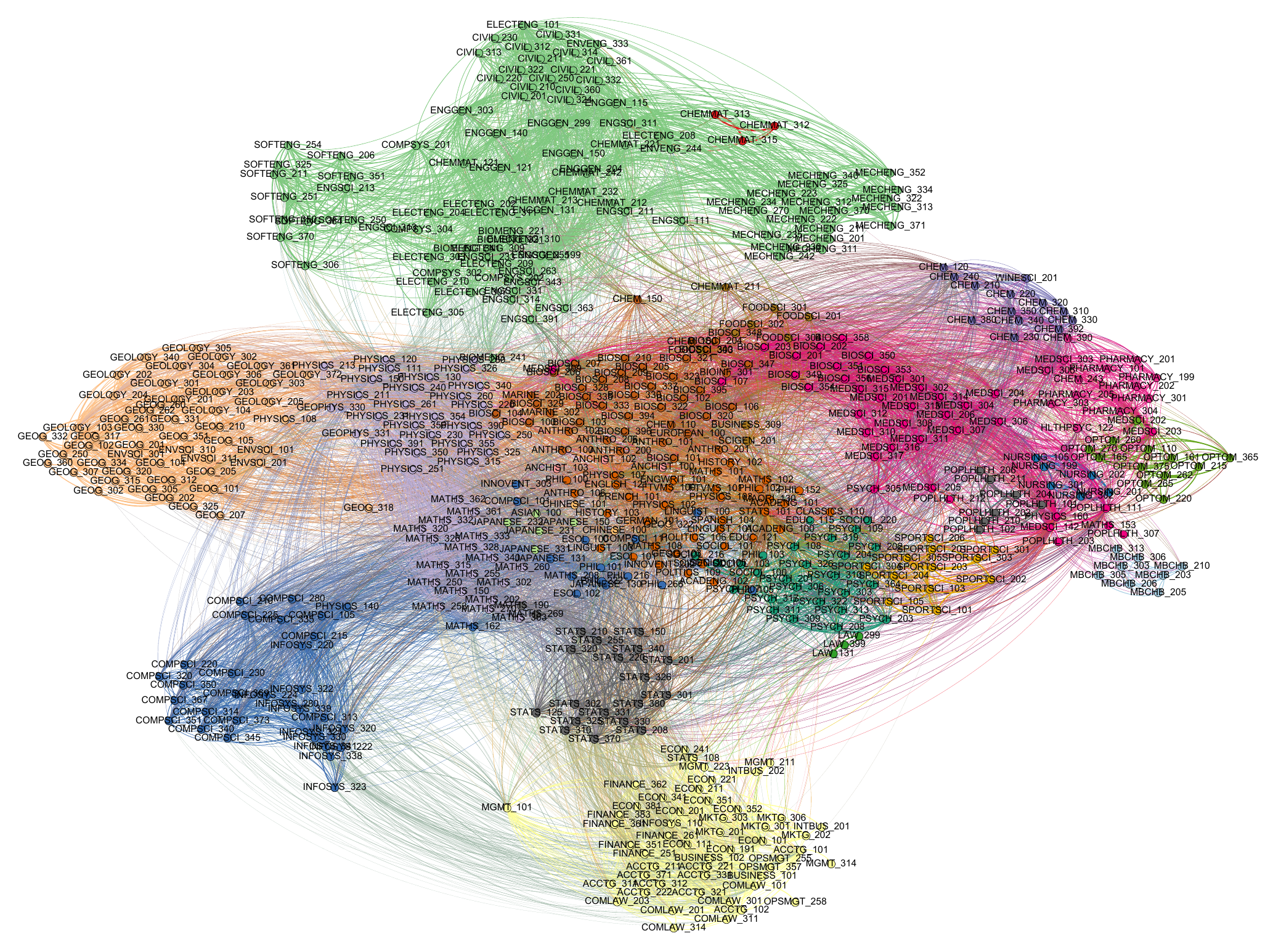}
\caption{\textbf{Student Course Network.\newline} 
A network representing the communities, or fields, of courses formed by students co-enrolling in course at the University of Auckland. Each node represents a course offered by the university, while links between nodes indicate instances where students took those two courses together within their undergraduate degree. Communities were revealed in a two step process. Firstly, edges were filtered so only those with an RCP value over 1 were included. Secondly, the Map Equation software package \cite{Edler_2017} was used to highlight the underlying fields. The revealed fields are labeled in Fig \ref{fig:Fig3}, and represent the various academic fields that students in our sample were enrolled in.}
\label{fig:Fig2}
\end{figure*}

\begin{figure*}[!ht]
\centering
\includegraphics[width = 13cm]{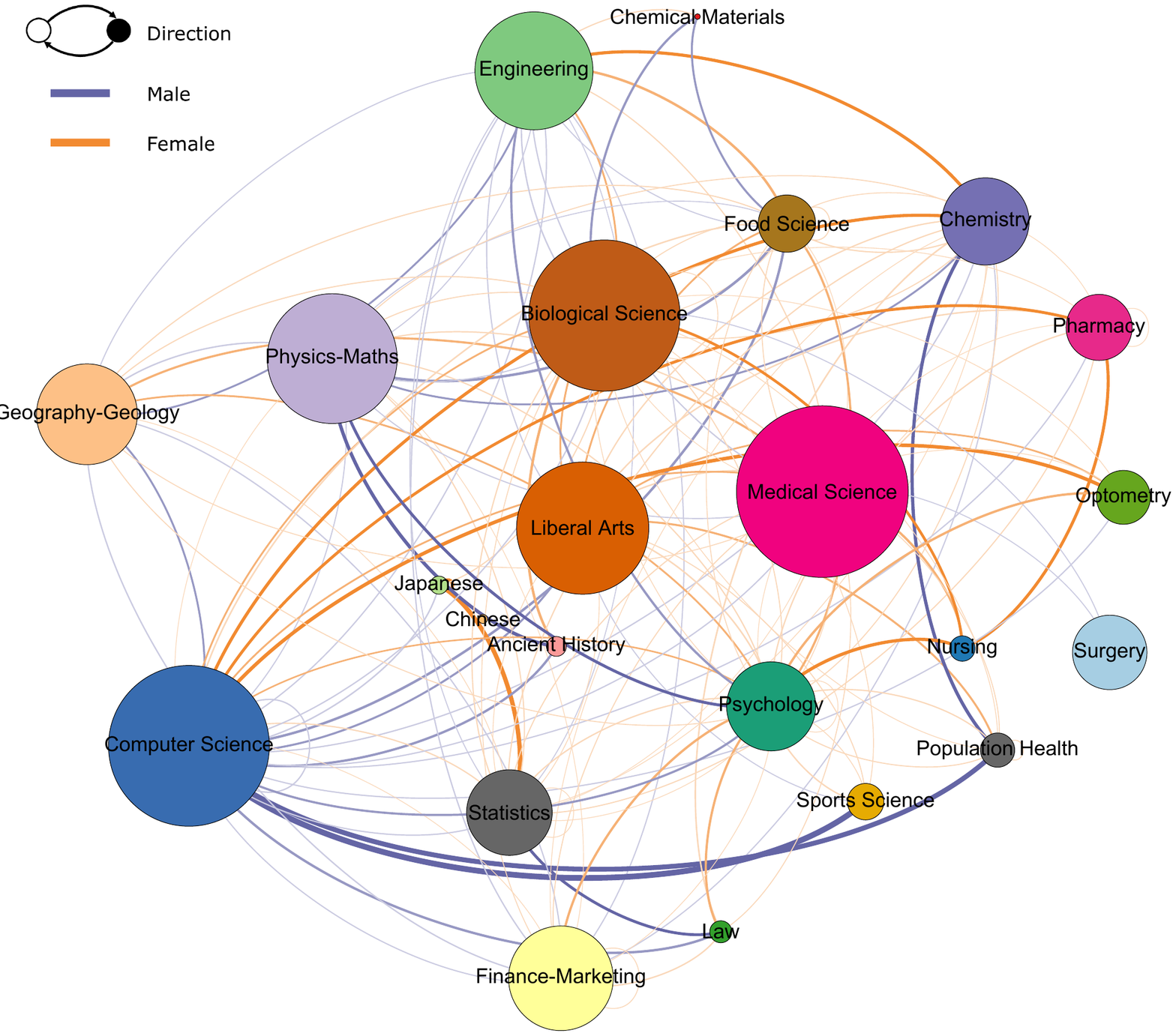}
\caption{\textbf{Course Community Network.\newline} 
The above directed network represents the network seen in Fig~\ref{fig:Fig2}, only the links within communities (i.e. links between courses belonging to the same community) and between communities have been split by gender and aggregated. Odds ratios comparing the likelihood of a female student taking a course in community$_i$ and community$_j$ were formulated, with the resulting values used as edge weights. The communities were labeled based on the range of courses that it is comprised of. Edges where female students were more likely to take a course in community$_i$ and community$_j$ are coloured blue, while edges where male students were more likely to take a course in community$_i$ and community$_j$ are coloured red. When considering the flow between a pair of nodes connected by two edges, the direction of flow is outward following the link in a clockwise direction. The network shows that transverse movements from fields such as computer science and physics-maths to other domains tend to be female dominated, whilst movements into these fields are more male dominated.}
\label{fig:Fig3}
\end{figure*}

\begin{figure*}[!ht]
\centering
\includegraphics[width = 13cm]{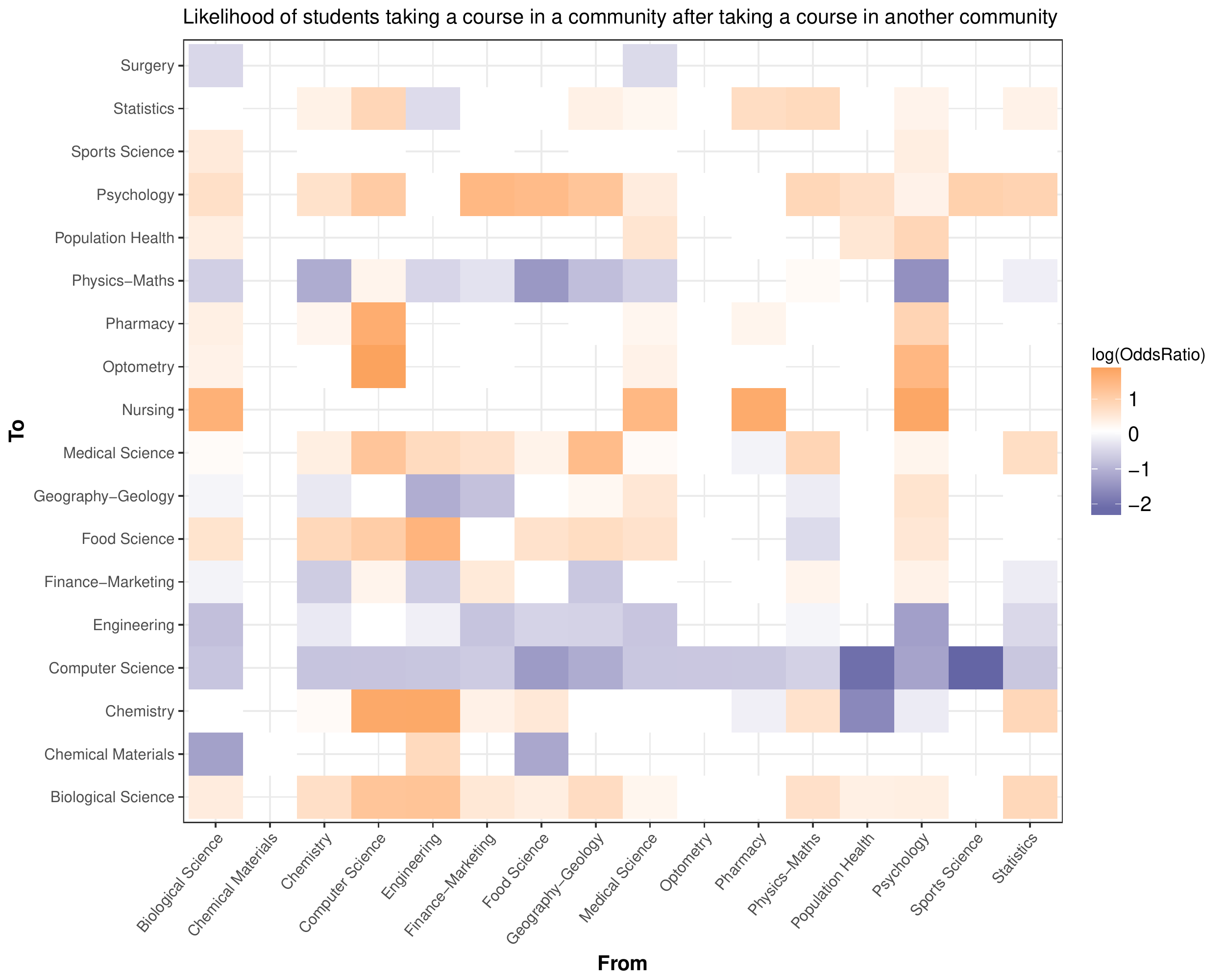}
\caption{\textbf{Student Course Heat Map.\newline} 
The above heat map represents the same underlying data as that which is used in Fig \ref{fig:Fig3}. The heat map makes clear the gender differences in the likelihood of students moving from one community to another. Orange areas represent instances where female students were more likely to take a course in community$_i$ after taking a course in community$_j$. Purple areas indicate male students were more likely to take a course in community$_i$ after taking a course in community$_j$. Areas that are white or empty indicate no significant relationship. Male students were consistently more likely to take courses in Computer Science and Physics-Maths after taking courses in each other community. Female students tended to be more likely to take courses in life science subjects (e.g., Biological Science and Psychology). }
\label{fig:Fig4}
\end{figure*}

\begin{figure*}[!ht]
\centering
\includegraphics[width = 13cm]{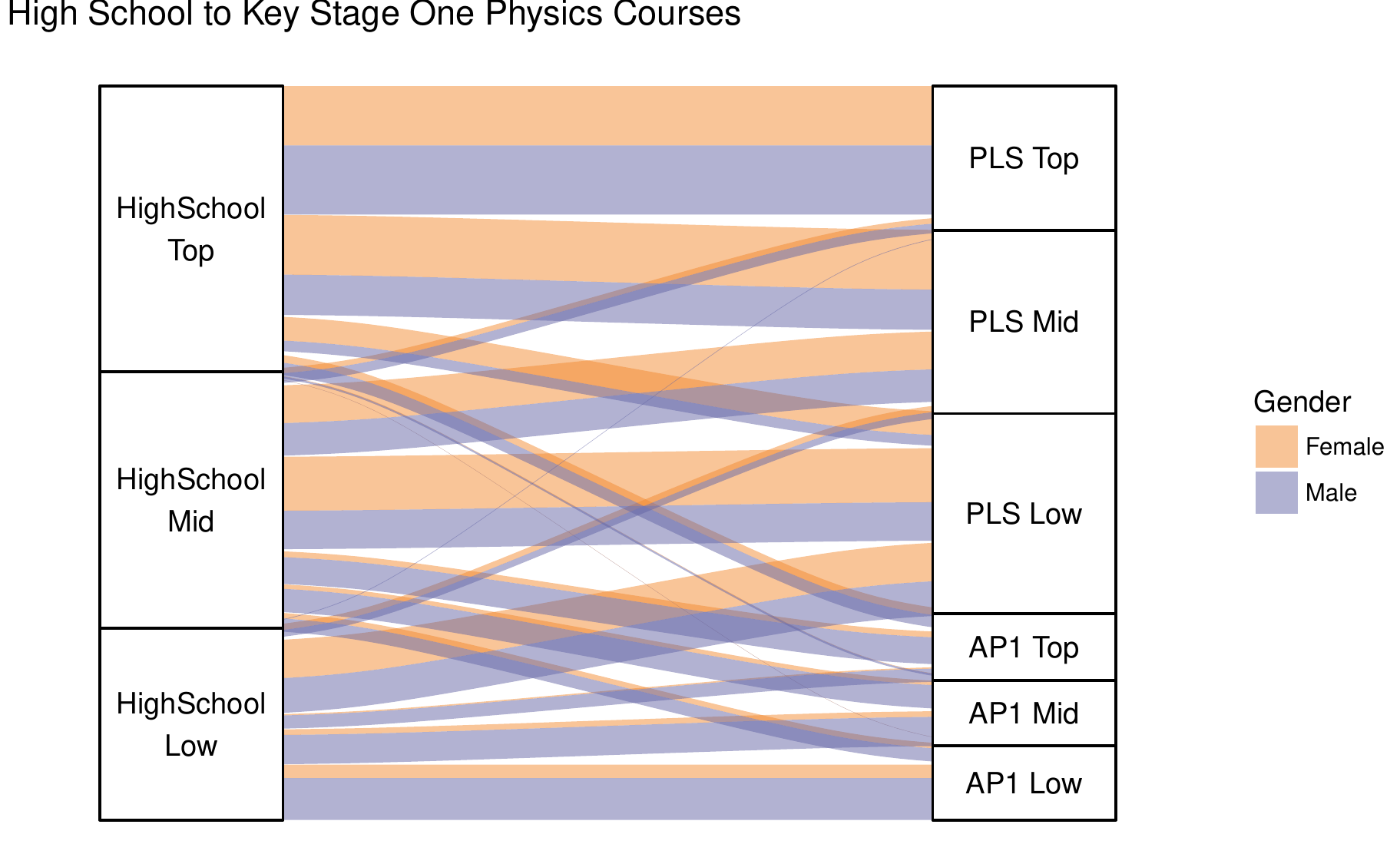}
\caption{\textbf{Student Progression Alluvial.\newline}
An alluvial plot showing the progression of students from high school to university physics split by achievement bands. PLS (Physics for Life Sciences) and AP1 (Advancing Physics 1) represent the two main groups of physics students in our data. As shown in the alluvial plot, PLS is more popular than AP1, especially at the intersection of top achievers and female students.}
\label{fig:Fig5} 
\end{figure*}

\begin{figure*}[!ht]
\centering
\includegraphics[width = 13cm]{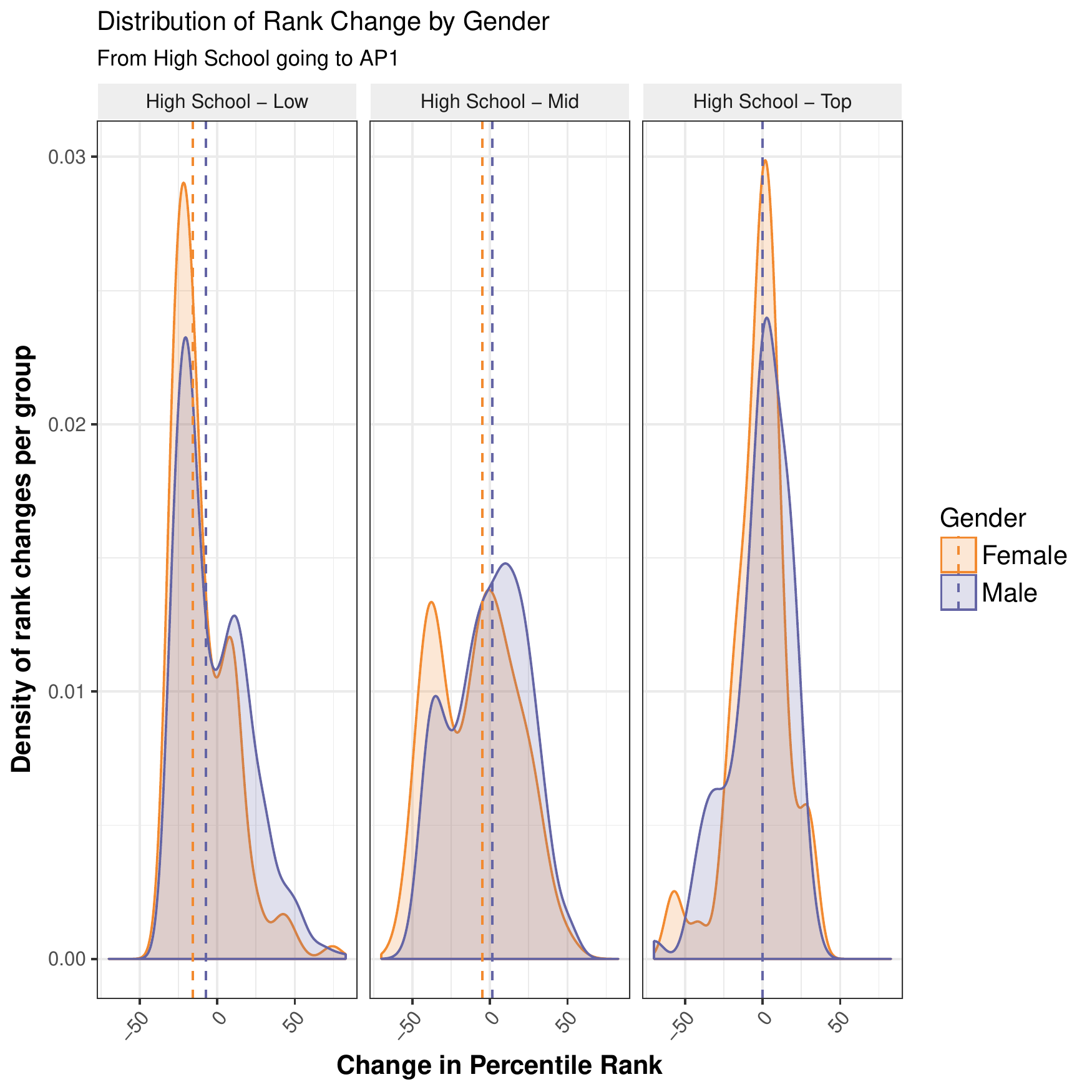}
\caption{\textbf{Distribution of Rank Change by Gender and High School Achievement Group.\newline} 
The above density plots show the distribution of rank change going from high school to Advancing Physics 1 (AP1). Purple represents the distribution of rank changes for male students, while orange represents female students. The dotted vertical line show the median rank change per group. On average, low achieving female students went down 6 ranks compared to their male counterparts, while middle achieving female students went down 9 ranks relative to their male counterparts. There was no significant gender difference in rank change for top achievers.}
\label{fig:Fig6}
\end{figure*}

\end{document}